\documentclass[times,sort&compress,3p]{elsarticle}
\journal{Journal of Multivariate Analysis}
\usepackage[labelfont=bf]{caption}

\usepackage{amsmath,amsfonts,amssymb,amsthm,color,epsfig,graphicx,hyperref,url}

\usepackage{booktabs}
\usepackage{longtable}
\usepackage{array}
\usepackage{multirow}
\usepackage{wrapfig}
\usepackage{float}
\usepackage{colortbl}
\usepackage{tabu}
\usepackage{threeparttable}
\usepackage{threeparttablex}
\usepackage[normalem]{ulem}
\usepackage{makecell}
\usepackage{mathtools}

\theoremstyle{plain}
\newtheorem{theorem}{Theorem}

\newtheorem{lemma}{Lemma}

\theoremstyle{definition}

\begin{document}

\begin{frontmatter}

\title{Sequential estimation of Spearman rank correlation using Hermite series estimators}

\author[mymainaddress]{Michael Stephanou\corref{mycorrespondingauthor}}

\author[mymainaddress,mysecondaryaddress]{Melvin Varughese}

\address[mymainaddress]{Department of Statistical Sciences, University of Cape Town, Cape Town, South Africa}
\address[mysecondaryaddress]{School of Mathematics and Statistics, University of Western Australia, Perth, Australia}

\cortext[mycorrespondingauthor]{Corresponding author. Email address: \url{michael.stephanou@gmail.com}}

\begin{abstract}
In this article we describe a new Hermite series based sequential estimator for the Spearman rank correlation coefficient and provide algorithms applicable in both the stationary and non-stationary settings. To treat the non-stationary setting, we introduce a novel, exponentially weighted estimator for the Spearman rank correlation, which allows the local nonparametric correlation of a bivariate data stream to be tracked. To the best of our knowledge this is the first algorithm to be proposed for estimating a time varying Spearman rank correlation that does not rely on a moving window approach. We explore the practical effectiveness of the Hermite series based estimators through real data and simulation studies demonstrating good practical performance. The simulation studies in particular reveal competitive performance compared to an existing algorithm. The potential applications of this work are manifold. The Hermite series based Spearman rank correlation estimator can be applied to fast and robust online calculation of correlation which may vary over time. Possible machine learning applications include, amongst others, fast feature selection and hierarchical clustering on massive data sets. \let\thefootnote\relax\footnote{\copyright\, 2021. This manuscript version is made available under the CC-BY-NC-ND 4.0 license \url{http://creativecommons.org/licenses/by-nc-nd/4.0/}}
\end{abstract}

\begin{keyword}
Hermite series estimators \sep
Incremental estimation \sep
Nonparametric correlation \sep
$O(1)$ update algorithm \sep
Online estimation \sep 
Sequential estimation \sep
Spearman correlation coefficient
\MSC[2020] 62L12\sep  62G05
\end{keyword}

\end{frontmatter}

\section{Introduction}

The statistical analysis of streaming data and one-pass analysis of massive data sets has become highly relevant in recent times. These settings necessitate online algorithms that are able to process observations sequentially, where ideally the time taken and memory used do not grow with the number of previous observations, i.e., $O(1)$ update time and memory requirements. In the univariate setting, certain statistical properties naturally lend themselves to efficient, sequential calculation such as the mean and variance \cite{welford,neely1966comparison,west1979updating,chan1982updating} and higher order moments \cite{bennett2009numerically,pebay2016numerically}. Similar developments include incremental formulae for cumulants up to fourth order \cite{amblard1995adaptive,dembele1998recursive}. In addition, algorithms have been introduced for sequential estimation of probability densities (see Chapters 4 and 5 of \cite{greblicki2008nonparametric} and Chapter 7 of \cite{devroye1985nonparametric} for a discussion of recursive kernel estimators and \cite{slaoui2019recursive} for recursive estimators based on Bernstein polynomials), cumulative probabilities \cite{slaoui2014stochastic,jmaei2017recursive,stephanou2017sequential} and quantiles \cite{P2, extP2First,extP2Second, P2ewma,ewsa,stephanou2017sequential,yazidi2017multiplicative, hammer2019tracking, hammer2019new, tiwari2019technique}. In the context of the statistical analysis of bivariate streaming data and one-pass analysis of massive bivariate data sets, certain quantities again easily extend to online calculation, such as the Pearson product-moment correlation coefficient. By contrast, online algorithms for nonparametric measures of concordance such as the Spearman rank correlation coefficient (Spearman Rho) and Kendall rank correlation coefficient (Kendall Tau) have only recently been proposed \cite{xiao2019novel}. These nonparametric correlation measures are suitable for all monotonic relationships and not just linear relationships as in the case of the Pearson correlation coefficient \cite{gibbons2010nonparametric}. In addition, these nonparametric correlation measures are more robust than the Pearson correlation estimator \cite{croux2010influence}. Applications of nonparametric correlation measures include eliciting relationships for financial instruments \cite{alanyali2013quantifying}, amongst others.
We propose a novel approach to the sequential estimation of the most popular nonparametric correlation measure, Spearman rank correlation, based on bivariate Hermite series density estimators and Hermite series based cumulative distribution function (CDF) estimators \cite{stephanou2017sequential,stephanou2020properties}. We leverage a key advantage of these estimators, namely their ability to maintain sequential estimates of the full density and full distribution function respectively. The central idea is that we can utilize the Hermite series CDF estimator and the bivariate Hermite series density estimator together with a large sample definition of the Spearman rank correlation estimator to furnish online estimates of the Spearman rank correlation. These estimates can be updated in constant, i.e., $O(1)$ time and require only a small and fixed amount of memory ($O(1)$ memory requirements with respect to number of observations). By comparison, the standard estimator of Spearman rank correlation requires first sorting all observations to determine ranks implying an average time complexity of $O(n\log{n})$ for $n$ observations. In addition, a na\"ive approach to updating a Spearman rank correlation estimate with a new observation would change all ranks (in general an operation with worse than constant time complexity) and necessitate a recalculation over the history of previous observations ($O(n)$ time complexity).
Our algorithms are useful in the stationary sequential estimation setting (i.i.d. observations from a bivariate distribution for example) as well as one-pass batch estimation in the setting of massive data sets. In the i.i.d. observation case, we are able to provide asymptotic guarantees on the rate of convergence in mean of this estimator to the large sample Spearman rank correlation estimate. The Hermite series based Spearman rank correlation estimation algorithm can also be modified to estimate the Spearman rank correlation for non-stationary bivariate data streams. To treat the case of sequential estimation in the non-stationary setting, we introduce a novel, exponentially weighted estimator for the Spearman rank correlation, which allows the local nonparametric correlation of a bivariate data stream to be tracked. To the best of our knowledge this is the first algorithm to be proposed for estimating a time varying Spearman rank correlation that does not rely on a moving window approach.
The rest of the article is organized as follows: in Section \ref{background} we review some relevant background on the Spearman rank correlation coefficient. In addition, we briefly review the bivariate Hermite series density estimator and the Hermite series distribution function estimator. In Section \ref{hermiteAndRank} we link the Hermite series based density and CDF estimators to the Spearman rank correlation coefficient. In Section \ref{stationaryAlgo} we present an algorithm for calculating the Spearman rank correlation coefficient applicable to the stationary sequential setting. In Section \ref{nonstationaryAlgo} we present an algorithm suitable for non-stationary data streams, based on an exponentially weighted Spearman rank correlation estimator. We provide the rate of convergence in mean of the Hermite series based Spearman rank correlation estimator applicable to the stationary setting for i.i.d data streams in Section \ref{theoreticalMAE}. In Section \ref{varianceEWEst} we investigate the variance (and hence standard error) properties of the exponentially weighted Spearman correlation estimator. In Section \ref{simstudies} we present simulation studies which demonstrate the effectiveness of our algorithms in practice both in the stationary and non-stationary settings. In the stationary setting, we demonstrate that our algorithm is competitive with an existing algorithm. In addition, we present an application of the non-stationary Spearman correlation estimation algorithm to real data in the form of streaming forex data in Section \ref{realResults}. We conclude in Section \ref{conclusion}. Proofs and further technical details are collected in \ref{appendix_proofs} and \ref{ewpearson}. An \textsf{R} \cite{rcore} package implementing the algorithms described in this article is available online \cite{hermiter}.

\section{Background}\label{background}

\subsection{Measures of Correlation}\label{measuresBackground}
In this article we focus our attention on the most popular measure of nonparametric correlation, namely the Spearman rank correlation coefficient. Suppose we have a sample of $n$ observations drawn from a continuous bivariate probability distribution, $F(x,y)$, with probability density function $f(x,y)$, i.e., $(\mathbf{x}_{i}$,$\mathbf{y}_{i}) \sim f(x,y)$. The Spearman rank correlation coefficient is defined as the sample Pearson product-moment correlation of the ranks of the observations,  $(r_{i}$,$s_{i}), \, i \in \{1,\dots, n\}$,
\begin{equation}
	R  = \frac{\sum_{i=1}^{n} (r_{i}-\bar{R})(s_{i}-\bar{S})}{(\sum_{i=1}^{n} (r_{i}-\bar{R})^{2}\sum_{i=1}^{n} (s_{i}-\bar{S})^{2})^{1/2}},\label{spearmanRank}
\end{equation}
where $\bar{R}=(1/n)\sum_{i=1}^{n} r_{i}$ and  $\bar{S}=(1/n)\sum_{i=1}^{n} s_{i}$ are the sample means of the ranks. The coefficient $R$ does not directly have a population analog. This is due to the fact that if we assume the marginal distributions of the random variables are continuous, the values which the random variables can take cannot be enumerated and ranked. However, we can define a constant which is a natural estimand as the sample size, $n \to \infty$, namely the grade correlation, $\rho(F^{(1)}(X), F^{(2)}(Y)),$ where $F^{(1)}(X), F^{(2)}(Y)$ are the marginal cumulative distribution functions and $\rho(x,y)$ is the Pearson product-moment correlation. In particular, \eqref{spearmanRank} is a Fisher consistent estimator of the grade correlation. We can also define the grade correlation as the constant for which $R$ is an unbiased estimator in large samples \cite{gibbons2010nonparametric},
\begin{equation*}
	lim_{n\to\infty} \mbox{E}(R) = \rho(F^{(1)}(X), F^{(2)}(Y)).
\end{equation*}
The grade correlation can be simplified to (using the exposition in \cite{gibbons2010nonparametric}),
 \begin{equation}
      \rho(F^{(1)}(X), F^{(2)}(Y)) = 12\int\int(F^{(1)}(x) - 1/2)(F^{(2)}(y) - 1/2)f(x,y)dx dy, \label{gradecorrel}
\end{equation}
an expression that will be useful in the developments in Section \ref{hermiteAndRank} below where we define an estimator for the grade correlation based on Hermite series bivariate density estimators and Hermite series based distribution function estimators.

\subsection{Hermite Series Estimators for Spearman Rank Correlation Estimation}\label{hermiteAndRank}

Hermite series estimators for univariate probability density functions have the following form \cite{schwartz1967estimation,convergence1,convergence2, greblicki1985pointwise,liebscher1990hermite}:
\begin{equation}
	\hat{f}_{N,n}(x) = \sum_{k=0}^{N} \hat{a}_{k}^{(n)} h_{k}(x), \quad \hat{a}_{k}^{(n)} = \frac{1}{n}\sum_{i=1}^{n} h_{k}(x_{i}), \quad k \in \{0,\dots,N\}, \label{HermiteSeriesProbEst}
\end{equation}
where $h_{k}=\left(2^{k}k!\sqrt{\pi}\right)^{-\frac{1}{2}} e^{-\frac{x^2}{2}} H_{k}(x), \, k \in \{0,\dots,N\}$, are the normalized Hermite functions defined from the Hermite polynomials, $H_k(x)=(-1)^k e^{x^2} (d^k / dx^k)e^{-x^2},\, k \in \{0,\dots,N\}$. The bivariate Hermite series probability density function estimator is given by:
\begin{equation}
	\hat{f}_{N_{1},N_{2},n}(x,y) = \sum_{k=0}^{N_{1}} \sum_{j=0}^{N_{2}} \hat{A}_{kj}^{(n)} h_{k}(x)h_{j}(y), \quad \hat{A}_{kj}^{(n)} = \frac{1}{n}\sum_{i=1}^{n} h_{k}(x_{i})h_{j}(y_{i}), \quad k \in \{0,\dots,N_{1}\}, \, j \in \{0,\dots,N_{2}\}, \label{HermiteSeriesProbEstBivariate}
\end{equation}
where $(\mathbf{x_{i}}, \mathbf{y_{i}}) \sim f(x,y)$.
Theoretical properties of the univariate Hermite series density estimator easily generalize to the multivariate case as discussed in \cite{schwartz1967estimation} and \cite{convergence1}. In \cite{stephanou2017sequential, stephanou2020properties} the following univariate distribution function estimator was studied, based on (\ref{HermiteSeriesProbEst}), for densities with support on the full real line:
\begin{equation}
\label{HermiteSeriesDistroEst}
	\hat{F}_{N,n}(x) = \int_{-\infty}^{x} \hat{f}_{N,n}(t) dt  = \sum_{k=0}^{N} \hat{a}_{k}^{(n)} \int_{-\infty}^{x} h_{k}(t) dt. 
\end{equation}
In the context of online estimation of the Spearman rank correlation coefficient we make use of the expression \eqref{gradecorrel}, the Hermite series bivariate density function estimator \eqref{HermiteSeriesProbEstBivariate} and the univariate Hermite series based distribution function estimator \eqref{HermiteSeriesDistroEst} to define:
 \begin{equation}
     \hat{R}_{N_{1},N_{2},N_{3},N_{4}}= 12\int\int(\hat{F}_{N_{1}}^{(1)}(x) - 1/2)(\hat{F}_{N_{2}}^{(2)}(y) - 1/2)\hat{f}_{N_{3} N_{4}}(x,y) dx dy. \label{hermiteSpearmansEst}
\end{equation}
As discussed in Section \ref{measuresBackground}, the grade correlation is that constant for which \eqref{spearmanRank} is an asymptotically unbiased estimator. Given that the online streaming and massive data set scenarios are large sample situations, the above estimator is natural. Note that in principle, $N_{1}$, $N_{2}$, $N_{3}$ and $N_{4}$ can all take on distinct values. However, for simplicity of the explication of the algorithms below (along with certain computational advantages), we set  $N_{1}=N_{2}=N_{3}=N_{4}=N$,
 \begin{equation}
     \hat{R}_{N}= 12\int\int(\hat{F}_{N}^{(1)}(x) - 1/2)(\hat{F}_{N}^{(2)}(y) - 1/2)\hat{f}_{NN}(x,y) dx dy. \label{hermiteSpearEstFull}
\end{equation}
For computational efficiency it is advantageous to phrase the estimator above in terms of linear algebra operations:
\begin{align}
     \hat{R}_{N} &= 12 (\mathbf{\hat{a}}_{(1)}^{(n)})^{\intercal} \mathbf{W}^{\intercal}\mathbf{\hat{A}}^{(n)}\mathbf{W}\mathbf{\hat{a}}_{(2)}^{(n)} \nonumber \\ &- 6 (\mathbf{\hat{a}}_{(1)}^{(n)})^{\intercal} \mathbf{W}^{\intercal} \mathbf{\hat{A}}^{(n)} \mathbf{z} - 6 \mathbf{z}^{\intercal} \mathbf{\hat{A}}^{(n)} \mathbf{W}\mathbf{\hat{a}}_{(2)}^{(n)} + 3  \mathbf{z}^{\intercal} \mathbf{\hat{A}}^{(n)} \mathbf{z} \mathord{,}\label{hermiteSpearmansEstMat} 
\end{align}
where the matrix $(\mathbf{W})_{kl} = \int_{-\infty}^{\infty}h_{k}(u)  \int_{-\infty}^{u} h_{l}(v) dv du$ and the vector $(\mathbf{z})_{k} = \int_{-\infty}^{\infty} h_k(u) du$. The vector $\mathbf{\hat{a}}_{(1)}^{(n)}$ is the vector of coefficients $(\mathbf{\hat{a}}_{(1)}^{(n)})_{k}$ associated with the Hermite series distribution function estimator $\hat{F}_{N}^{(1)}(x)$ and $\mathbf{\hat{a}}_{(2)}^{(n)}$ is the vector of coefficients $(\mathbf{\hat{a}}_{(2)}^{(n)})_{k}$ associated with the Hermite series distribution function estimator $\hat{F}_{N}^{(2)}(y)$. The vectors $\mathbf{\hat{a}}_{(1)}^{(n)}, \mathbf{\hat{a}}_{(2)}^{(n)} $ are as defined in equation \eqref{HermiteSeriesProbEst}. Sequential, but ultimately equivalent, versions of these estimators are presented in equations \eqref{hermite_cdf_stat_1} and \eqref{hermite_cdf_stat_2} for stationary data. In the non-stationary case we can utilize exponentially weighted estimates instead as per equations \eqref{hermite_cdf_exp_1} and \eqref{hermite_cdf_exp_2}. The matrix $\mathbf{\hat{A}}^{(n)}$ is the matrix of coefficients $(\mathbf{\hat{A}}^{(n)})_{kl}$ associated with the Hermite series bivariate density function estimator $\hat{f}_{NN}(x,y)$ as defined in equation \eqref{HermiteSeriesProbEstBivariate}. A sequential, but equivalent estimator is provided in \eqref{hermite_pdf_stat_1} for stationary data. An exponentially weighted estimator appropriate to non-stationary data is provided in equation  \eqref{hermite_pdf_exp}. For all matrices and vectors thus defined,  $k,l \in \{0, \dots, N\}$. Note that these integrals can be evaluated numerically once and the values stored for rapid calculation. In the next section we utilize this estimator in order to define algorithms for sequential (online) Spearman rank correlation estimation in both stationary and non-stationary data settings.

\section{Sequential Spearman Rank Correlation Estimation}\label{seqSpearmanAlgoStat}

\subsection{Sequential Analysis of Stationary Data} \label{stationaryAlgo}

In this section, we propose an algorithm for sequential (online) estimation of the Spearman rank correlation coefficient in the setting of stationary data, namely stationary streaming data and massive data sets. \\

\noindent {\bf Algorithm 1}

\begin{enumerate}
	\item For each observation from the data stream $(\mathbf{x}_{i},\mathbf{y}_{i})$, where $i \in \{1, \dots, n\}$, apply the update rules:
	\begin{equation}
                     \mathbf{\hat{a}}_{(1)}^{(1)} =   \mathbf{h} (\mathbf{x}_{1}), \quad \mathbf{\hat{a}}_{(1)}^{(i)} = \frac{1}{i} \left[(i-1)\mathbf{\hat{a}}_{(1)}^{(i-1)} + \mathbf{h} (\mathbf{x}_{i})\right], \, i \in \{2,\dots,n\},\label{hermite_cdf_stat_1}
	\end{equation}
	\begin{equation}
		\mathbf{\hat{a}}_{(2)}^{(1)} =   \mathbf{h} (\mathbf{y}_{1}), \quad
		\mathbf{\hat{a}}_{(2)}^{(i)} = \frac{1}{i} \left[(i-1)\mathbf{\hat{a}}_{(2)}^{(i-1)} + \mathbf{h} (\mathbf{y}_{i})\right],  \, i \in \{2,\dots,n\},\label{hermite_cdf_stat_2}
	\end{equation}
	\begin{equation}
		\mathbf{\hat{A}}^{(1)} = \mathbf{h} (\mathbf{x}_{1}) \otimes \mathbf{h} (\mathbf{y}_{1}),\quad 
		\mathbf{\hat{A}}^{(i)} = \frac{1}{i} \left[(i-1)\mathbf{\hat{A}}^{(i-1)} +\mathbf{h} (\mathbf{x}_{i}) \otimes \mathbf{h} (\mathbf{y}_{i})\right], \, i \in \{2,\dots,n\}.\label{hermite_pdf_stat_1}
	\end{equation}
	\item Plug the coefficients $\mathbf{\hat{a}}_{(1)}^{(i)}$, $\mathbf{\hat{a}}_{(2)}^{(i)}$ and $\mathbf{\hat{A}}^{(i)}$ into the expression \eqref{hermiteSpearmansEstMat} to obtain an updated online estimate of the Spearman rank correlation coefficient, $\hat{R}_{N}$.
\end{enumerate}

Here $\mathbf{h} (\mathbf{x}_{i})$ and $\mathbf{h} (\mathbf{y}_{i})$ are the vectors $h_{k} (\mathbf{x}_{i}), \, k \in \{0,\dots,N\}$ and $h_{l} (\mathbf{y}_{i}), l \in \{0,\dots,N\}$ respectively. For computational efficiency it is advantageous to calculate $ \mathbf{h} (\mathbf{x}_{i})$ and $\mathbf{h} (\mathbf{y}_{i})$ making use of the recurrence relation for the Hermite polynomials $H_{k+1}(x) = 2xH_{k}(x) - 2kH_{k-1}(x)$. Note that in the practical application of this algorithm, $N$ is fixed and does not depend on the number of observations $n$. This is different to the setup in Theorem \ref{mainTheoremMAE}, where we show that the estimator \eqref{hermiteSpearmansEst} is consistent in mean for $N(n)\to \infty, n \to \infty$ under certain conditions, providing comfort that the estimator is sensible in an asymptotic sense. For fixed $N$, in the large sample limit where the number of observations, $n \to \infty$, we learn in the proof of Theorem \ref{mainTheoremMAE} that the asymptotic bound on the MAE of the estimator \eqref{hermiteSpearmansEst} decreases with increasing $N$. Thus, in the massive data set and streaming scenarios, $N$ would be set as large as possible subject to computational constraints (larger $N$ implies more costly numerical calculations). Indeed, the simulation studies in Section \ref{simstudies_stat} support this assertion. We demonstrate through extensive simulation studies that for the bivariate normal distribution across a range of correlation parameters, a value of $N=20$ is sufficiently large and gives good results. This value can therefore be used as a starting point for similar distributions.  The computational cost of updating the coefficients, $\mathbf{\hat{a}}_{(1)}^{(i)}$, $\mathbf{\hat{a}}_{(2)}^{(i)}$ and $\mathbf{\hat{A}}^{(i)}$, as above is manifestly constant ($O(1)$). In addition, since the expression $\eqref{hermiteSpearmansEstMat}$ has no explicit dependence on the observations, the time complexity of updating the Spearman correlation coefficient is also $O(1)$ with respect to the number of previous observations. Finally, it is clear that the space/memory complexity is $O(1)$ with respect to the number of previous observations since the number of coefficients to be stored and updated is fixed.

\subsection{Sequential Analysis of Non-Stationary Data} \label{nonstationaryAlgo}

In this section we describe an algorithm for tracking the Spearman rank correlation coefficient for a dynamically varying (non-stationary) data stream. To the best of our knowledge the algorithm we present below is the only online algorithm for the Spearman rank correlation coefficient applicable to non-stationary data streams that does not rely on maintaining a moving/sliding window of previous observations. Our approach is based on using an exponentially weighted moving average version of the Hermite series coefficients. The parameter $\lambda$ in the algorithm below controls the weighting of new observations (and controls how rapidly the weights of older observations decrease). This weighting scheme allows the local nonparametric correlation of a bivariate data stream to be tracked. \\

\noindent {\bf Algorithm 2}

\begin{enumerate}
	\item For each observation from the data stream $(\mathbf{x}_{i},\mathbf{y}_{i})$, where $i \in \{1, \dots, n\}$, apply the update rules:
	\begin{equation}
		\mathbf{\hat{a}}_{(1)}^{(1)} =   \mathbf{h} (\mathbf{x}_{1}), \quad
		\mathbf{\hat{a}}_{(1)}^{(i)} =  (1-\lambda) \mathbf{\hat{a}}_{(1)}^{(i-1)} + \lambda \mathbf{h}  (\mathbf{x}_{i}),  \, i \in \{2,\dots,n\}, \label{hermite_cdf_exp_1}
	\end{equation}
	\begin{equation}
		\mathbf{\hat{a}}_{(2)}^{(1)} =   \mathbf{h} (\mathbf{y}_{1}), \quad
		\mathbf{\hat{a}}_{(2)}^{(i)} = (1-\lambda) \mathbf{\hat{a}}_{(2)}^{(i-1)} + \lambda \mathbf{h}  (\mathbf{y}_{i}),   \, i \in \{2,\dots,n\}, \label{hermite_cdf_exp_2}
	\end{equation}
	\begin{equation}
		\mathbf{\hat{A}}^{(1)} = \mathbf{h} (\mathbf{x}_{1}) \otimes \mathbf{h} (\mathbf{y}_{1}),\quad
		\mathbf{\hat{A}}^{(i)} = (1-\lambda)\mathbf{\hat{A}}^{(i-1)} + \lambda \mathbf{h}  (\mathbf{x}_{i}) \otimes \mathbf{h} (\mathbf{y}_{i}), \, i \in \{2,\dots,n\}. \label{hermite_pdf_exp}
	\end{equation}
	\item Plug the coefficients $\mathbf{\hat{a}}_{(1)}^{(i)}$, $\mathbf{\hat{a}}_{(2)}^{(i)}$ and $\mathbf{\hat{A}}^{(i)}$ into the expression \eqref{hermiteSpearmansEstMat} to obtain an updated online estimate of the Spearman rank correlation coefficient, $\hat{R}_{N}$.
\end{enumerate}

Note that in the algorithm above, $N, \lambda$ are both fixed and do not depend on the number of observations $n$. Again the time complexity of updating the coefficients, $\mathbf{\hat{a}}_{(1)}^{(i)}$, $\mathbf{\hat{a}}_{(2)}^{(i)}$ and $\mathbf{\hat{A}}^{(i)}$, is manifestly constant ($O(1)$) as is the time and memory/space complexity of updating the Spearman correlation coefficient.

\subsection{Guidance in selecting $\lambda$ and $N$} \label{lam_N_selection}

In selecting $\lambda$, there are two balancing considerations, namely the responsiveness of the Hermite series based Spearman correlation estimator to changes in the non-stationary streaming data being analyzed and the variance in the estimates. We expect that the higher the value of $\lambda$, the more responsive the estimator, but that this responsiveness comes at the cost of increased variance. We can make both of these considerations more concrete. In terms of responsiveness of the estimator, $\lambda$ essentially determines an effective window size for past observations to include in the estimate. This can be understood by noting that the weighting factors of terms comprising the Hermite series coefficient estimates decrease geometrically with increasing observation age. Beyond a certain observation age, the contribution to the estimates become negligible in practice (particularly considering that the individual terms in the Hermite series coefficient estimators are bounded). One commonly used relation between simple moving averages and exponentially weighted moving averages suggests setting $\lambda = 2/(w+1)$, where $w$ is the moving window size of the simple moving average, to achieve an effective window size of $w$ for the exponentially weighted moving average. This relation is based on equating the average ``age"/lag of an observation in the moving window, $(w+1)/2$, and the average ``age"/lag of an observation in the exponentially weighted setting, $1/\lambda$. Using this relation, we see that one possible definition for the effective window size is $w = 2 / \lambda - 1$. While there are several possible definitions of the effective window size, all definitions imply that the larger the value of $\lambda$, the smaller the effective window size and the more responsive the estimator should be to recent changes and vice-versa.  In terms of the variance of the Hermite series based Spearman correlation estimator, we can make concrete statements in an i.i.d. scenario, as explored in Section \ref{varianceEWEst}. This analysis shows that the variance is $O(\lambda)$ and the standard error is $O(\lambda^{1/2})$. Using the results elucidated in Section \ref{varianceEWEst}, we obtain a simple rough guide to select $\lambda$ based on the standard error one is willing to tolerate, $\sigma_{tol}$, namely,
$
\lambda \lesssim 2\sigma_{tol}^{2}.
$
While these results are based on an i.i.d. scenario and the bivariate normal distribution in particular, we propose that they can be cautiously applied as a rule of thumb even in more general non-stationary scenarios. This simple guide also builds in the fact that given that the value of the Spearman correlation coefficient lies between $-1$ and $1$, it is reasonable to assume that a standard error of $\sigma_{tol} < 0.2$ is desirable.   Our results suggest that if one wants to control the standard error at say, $\sigma_{tol} = 0.1$, then $\lambda \lesssim 0.02$. This is supported by our simulation studies in Section \ref{simnonstat} and real data analyses in Section \ref{realResults} where we have found that smaller values of lambda (i.e. $0.002, 0.005,0.01$) give good results and are sensible starting points. The parameter $\lambda$ can also be varied over time in principle. Varying $\lambda$ over time in a data driven manner is a promising area for future research. In selecting $N$ along with $\lambda$ one needs to bear in mind that the effective window size may not be large enough to motivate setting a large $N$ value. Indeed, the effective window size does not grow with the number of observations, so the recommendation to set $N$ as large as possible subject to computational constraints as discussed in the stationary case (Section \ref{stationaryAlgo}) is no longer appropriate. We demonstrate in our simulation studies in Section \ref{simnonstat} that the performance of the estimator is not critically sensitive to the choice of $N$ however, and $N=10$ or $N=20$ should constitute reasonable starting points. In order to perform data-driven selection of the fixed $N$ parameter in stationary scenarios and fixed $N, \lambda$ parameters in non-stationary scenarios we recommend a grid search over $N$ and $N, \lambda$ values respectively minimizing a contemporaneous or predictive error function on an initial sample of the streaming data to be analyzed. This procedure is applied in Section \ref{realResults} with a loss function of predictive MAE to illustrate the approach. With sufficient initial data this grid search would ideally be combined with time-series cross-validation to establish generalizability. Note that in stationary scenarios, $N$ can be selected based on analysis of an initial sample in the knowledge that as $n \to \infty$, the accuracy will only improve.   

\section{Mean Absolute Error of Hermite based Estimator}\label{theoreticalMAE}

In this section, we present a theorem concerning the asymptotic convergence of the estimator \eqref{hermiteSpearmansEst} to the grade correlation in an i.i.d. setting as the number of observations $n \to \infty$. In particular, we prove convergence in mean and provide the rate. While these results do not directly apply to the fixed and finite $N_1, N_2, N_3, N_4$ scenario, they do give assurances that the asymptotic properties of the estimator \eqref{hermiteSpearmansEst} are sensible.

\begin{theorem} \label{mainTheoremMAE}

For a sample of $n$ i.i.d. bivariate observations $(\mathbf{x_{i}},\mathbf{y_{i}}) \sim f(x,y)$, assume:

\begin{enumerate}
	\item  $f^{(1)}(x) \in L_{2}$, $r_1 \geq 8$ derivatives of $f^{(1)}(x)$ exist, $(x-d/dx)^{r_1} f^{(1)}(x) \in L_{2}$ and $\mbox{E}|X|^{4/3}<\infty$, where $f^{(1)}(x)$ is the probability density function associated with the marginal cumulative distribution $F^{(1)}(x)$.
	\item  $f^{(2)}(y) \in L_{2}$, $r_2 \geq 8$ derivatives of $f^{(2)}(y)$ exist, $(y-d/dy)^{r_2} f^{(2)}(y) \in L_{2}$ and $\mbox{E}|Y|^{4/3}<\infty$, where $f^{(2)}(y)$ is the probability density function associated with the marginal cumulative distribution $F^{(2)}(y)$.
	\item $f(x,y) \in L_2$ and $(x-\partial_{x})^{r_3}(y-\partial_{y})^{r_3}f(x,y) \in L_{2}$, $r_3 \geq 8$.
	\item $N_{1}(n) = O(n^{2/(2r_1+1)})$, $N_{2}(n)= O(n^{2/(2r_2+1)})$ and $N_{3}(n) = N_{4}(n) = O(n^{1/(2r_3+1)})$ as $n \to \infty$.
\end{enumerate}

For $n\to \infty$:
\begin{align*}
\mbox{E}|\hat{R}_{N_{1},N_{2},N_{3},N_{4}} -  \rho(F^{(1)}(X), F^{(2)}(Y))| =  &O(n^{34/12(2r_{1}+1)+34/12(2r_{2}+1)-(r_3-2)/(2r_3+1)})+ O(n^{34/12(2r_{1}+1) -(r_2 -2)/(2r_2+1)})\nonumber \\
&+ O(n^{-(r_1 -2)/(2r_1+1)}) = o(1) \mbox{ uniformly in x,y}.\nonumber 
\end{align*}
 
\end{theorem}

See \ref{appendix_proofs} for the proof of Theorem \ref{mainTheoremMAE}. A few comments are in order. The theorem above only applies for $r_1, r_2, r_3$ fixed and finite as discussed in the proof in  \ref{appendix_proofs}. What does this imply for densities such as the normal distribution density? As demonstrated in \cite{stephanou2020properties}, all derivatives of $f(x)$ exist for the normal density and $(x-d/dx)^{r} f(x) \in L_{2}$ for all finite $r \geq 1$. In the context of Theorem \ref{mainTheoremMAE}, this case is directly relevant to the illustrative example where $f(x,y)$ is the bivariate normal distribution with $\rho=0$. In this case, the first three conditions are clearly satisfied for all finite $r_1, r_2, r_3 \geq 8$. Therefore, for a choice of $r_1, r_2, r_3 >> 1$, the asymptotic bound on the MAE is approximately $O(n^{-1/2})$, for sufficiently large $N(n)$ and $n$. It remains true that the larger the values of $N(n)$ and $n$ the lower the bound on the MAE. In general, while larger values of $r_1, r_2, r_3$ improve the asymptotic rate to a point, they affect the size of the coefficients in the asymptotic expansion and larger values of $N(n), n$ may be required before $\mbox{MAE} << 1$. It is also noteworthy that Theorem \ref{mainTheoremMAE} does not apply to distributions where the mean does not exist or the mean is not finite due to the moment conditions described in the first two conditions above.  

\section{Variance of Exponentially Weighted Hermite based Estimator} \label{varianceEWEst}

The estimator introduced in Section \ref{nonstationaryAlgo} based on exponentially weighted versions of the Hermite series coefficients should be applicable in the general non-stationary scenario. In the general case, we cannot necessarily obtain direct theoretical insights into the bias and variance of the Hermite series based Spearman rank correlation estimator (e.g. useful bounds etc.). We can however get insights into the variance of the estimator introduced in \ref{nonstationaryAlgo} in the i.i.d. scenario. In particular we can establish the relationship between the variance of $\hat{R}_N$ and $\lambda$ for fixed $N$ as $n \to \infty$.

\begin{theorem} \label{theoremVariance}

For fixed $N$, fixed $0< \lambda <1$ and $n\to \infty$ we have:

\begin{align*}
\mbox{Var}(\hat{R}_N) = &\left( \frac{\lambda}{2-\lambda} \right)[\sum_{r=0}^{N} (g^{(1)}_{r})^{2} \mbox{Var}(h_{r}(x)) + \sum_{s=0}^{N} (g^{(2)}_{s})^{2} \mbox{Var}(h_{s}(y)) + \sum_{u,v=0}^{N} (g^{(3)}_{uv})^{2} \mbox{Var}(h_{u}(x) h_{v}(y)) \\
&+ \sum_{r,s=0}^{N} g^{(1)}_{r} g^{(2)}_{s} \mbox{Cov}(h_{r}(x), h_{s}(y)) + \sum_{r,u,v=0}^{N} g^{(1)}_{r}g^{(3)}_{uv} \mbox{Cov}(h_{r}(x), h_{u}(x)h_{v}(y)) \\
&+ \sum_{s,u,v=0}^{N} g^{(2)}_{s}g^{(3)}_{uv} \mbox{Cov}(h_{s}(y), h_{u}(x)h_{v}(y))] + o\left( \lambda \right),
\end{align*}
where
\begin{align}
g^{(1)}_{r} &= \frac{\partial \hat{R}_N}{\partial (\hat{a}_{(1)})_{r}} |_{a_{(1)},a_{(2)},A} = 12 \sum_{l,m,o} W_{lr} A_{lm} W_{mo} (a_{(2)})_{o} -  6\sum_{l,m}W_{lr}A_{lm}z_{m}, \nonumber\\
g^{(2)}_{s} &= \frac{\partial \hat{R}_N}{\partial (\hat{a}_{(2)})_{s}} |_{a_{(1)},a_{(2)},A} = 12\sum_{k,l,m}(a_{(1)})_{k} W_{lk} A_{lm} W_{ms} - 6\sum_{k,m}z_{k} A_{km}W_{ms}, \nonumber
\end{align}
\begin{equation}
g^{(3)}_{uv} = \frac{\partial \hat{R}_N}{\partial (\hat{A}_{uv})} |_{a_{(1)},a_{(2)},A} = 12\sum_{k,o}(a_{(1)})_{k} W_{uk}W_{vo} (a_{(2)})_{o} - 6\sum_{k}(a_{(1)})_{k}W_{uk}z_{v} \nonumber\\ 
- 6\sum_{o}z_{u}W_{vo} (a_{(2)})_{o} +3z_{u} z_{v}. \nonumber
\end{equation}
Thus $\mbox{Var}(\hat{R}_N)= O(\lambda)$ and the standard error of $\hat{R}_N$ is $O(\lambda^{1/2})$.
\end{theorem}
See \ref{appendix_proofs} for a proof of Theorem \ref{theoremVariance}. As is fruitful in settings like bandwidth selection for kernel density estimation, we can obtain specific results for the normal distribution and use these as a rough guide to the behaviour of similar distributions (e.g. Silverman's rule of thumb). We begin by restating the result above for the bivariate normal distribution for $n \to \infty$ as:
\begin{equation}
\mbox{Var}(\hat{R}_N) = \left[ \frac{\lambda}{2-\lambda} \right] g_{N,\rho} + o\left( \lambda \right).\nonumber
\end{equation}

As noted in Section \ref{lam_N_selection}, the larger the value of $\lambda$, the more responsive the estimator will be to changes in the distribution of the data stream being analyzed. We can utilize the results for the normal distribution to provide a simple rough guide as to how large one can set $\lambda$ to increase responsiveness whilst staying within a chosen variance ``budget". These results are particularly relevant to non-stationary data streams which switch between different stationary regimes for example. Roughly speaking, the variance for sufficiently small $\lambda$ is approximately $\left[ \lambda / (2-\lambda) \right] g_{N,\rho}$ for distributions similar to the bivariate normal distribution with correlation parameter $\rho$ in the i.i.d. scenario. We have evaluated the values of $g_{N,\rho}$ numerically and determined that the maximum variance across a range of $N, \rho$ values is approximately $\left[\lambda / (2-\lambda) \right]$ (obtained at $\rho=0$).  Thus, $\lambda$ should be selected according to, $\lambda \lesssim (2\sigma_{tol}^{2})/(1+\sigma_{tol})^{2}$, in order to stay within the variance budget defined by $\sigma_{tol}^{2}$. In fact, given that one would reasonably set $\sigma_{tol}$ to a small value, e.g., $\sigma_{tol} \leq 0.2$, we can apply:
\begin{equation*}
	\lambda \lesssim 2\sigma_{tol}^{2},
\end{equation*}
furnishing a simple rough guide for selecting $\lambda$. These results may even be potentially applied in a weakly serially correlated scenario. This is an area for further research.

\section{Simulation Studies}\label{simstudies}

\subsection{Stationary Data}\label{simstudies_stat}

In this section, we evaluate the effectiveness of the proposed sequential Hermite series based Spearman rank correlation estimation algorithm and compare to the first and only existing algorithm known to the authors,  \cite{xiao2019novel}, which we will refer to as the count matrix based approach. The count matrix algorithm provides a neat and straightforward way to approximate the Spearman rank correlation coefficient for streaming data in a sequential manner. In essence, observations are placed in two-dimensional buckets, similar to constructing a two dimensional histogram with fixed cut-points. These cut-points are chosen to be quantiles of the univariate standard normal distribution. Upon the arrival of a new observation, the appropriate bucket's count is incremented. The associated matrix of observation counts forms an input to a tied-observation form of the Spearman rank correlation coefficient estimator, phrased in terms of linear algebra operations for computational efficiency. It is worth briefly noting here that the count matrix algorithm can only be applied in non-stationary scenarios using a moving window approach, where memory requirements grow linearly with chosen window size. By contrast, the Hermite series based algorithm can be easily extended to the non-stationary scenario (i.e. algorithm 2 in Section \ref{nonstationaryAlgo}) and has fixed memory requirements irrespective of the ``effective" window size implied by the weighting parameter $\lambda$. This is a distinct advantage. This implies that even if the accuracy of the algorithms was similar, the Hermite series based approach is still a valuable development.

The simulation study we conduct is as follows, we draw i.i.d. samples from the ubiquitous bivariate normal distribution for several sample sizes, $n$, at various values of the correlation parameter, $\rho$, and compare the accuracy of the sequential Hermite series based Spearman rank correlation estimation algorithm to the count matrix based algorithm. We utilize the following parameters for the mean vector and covariance matrix respectively, $\mu = (0,0)$, $\Sigma = (\sigma_{1}^{2}, \rho \sigma_{1}\sigma_{2}; \rho\sigma_{1}\sigma_{2},\sigma_{2}^{2})$ where $\sigma_{1}=1, \sigma_{2}=1$. For each sample size $n \in \{10^{4},5\times10^{4},10^{5}\}$ and each value of the correlation parameter, $\rho \in \{-0.75, -0.5, -0.25, 0.25,0.5, 0.75\}$, the following steps are repeated $m=10^3$ times for the Hermite series based Spearman rank correlation estimation algorithm:
\begin{enumerate}
\item Draw $n$ i.i.d. observations, $(x_{i},y_{i}), \, i \in \{1,\dots,n\}$ from the bivariate normal distribution with mean vector $\mu$ and correlation parameter $\rho$;
\item Iterate through the sample, updating the Hermite series based Spearman rank correlation estimate, $\hat{R}_{N}$, in accordance with algorithm 1 in Section \ref{stationaryAlgo}. The Spearman rank correlation is then recorded at the last observation in the sample, $i=n$, for the Hermite series based algorithm;
\item Calculate the exact Spearman correlation coefficient, $R$, for the sample;
\item Calculate the absolute error between the Hermite series based Spearman correlation estimate and the exact Spearman correlation coefficient.
\end{enumerate}
The MAE between the Hermite series based estimate and the exact Spearman rank correlation coefficient for a particular value of $\rho$ and sample size $n$ is then estimated through $\widehat{\mbox{MAE}}(\hat{R}_{N}) = (1/m) \sum_{j=1}^{m} |\hat{R}^{(j)}_{N} - R^{(j)}|$, where $j$ indexes a particular set of $n$ observations and $m=10^3$ is the number of repeated draws of $n$ observations. The exact same steps as above are repeated for the count matrix algorithm, where we iterate through each sample, updating the count matrix based Spearman rank correlation estimate in accordance with algorithm 2 in \cite{xiao2019novel}. 

In terms of the choice of $N$ for the Hermite series based algorithm, we selected a value consistent with our considerations in Section \ref{stationaryAlgo}. Given that our simulation studies are conducted for large samples, we would expect that the larger the value of $N$, the better the accuracy (as long as our samples are sufficiently large). A coarse grid search of $N$ values, up to a maximum of $N=20$, confirmed that the largest value, $N=20$ indeed minimized the average estimated MAE across all values of $\rho$ for all  sample sizes considered. The maximum value of $N$ for the grid search, $N=20$, was chosen as a value yielding very good computational performance in our context. We evaluated two choices of cut-points for the count matrix based algorithm, namely $c=20$ and $c=30$. The choices are motivated as follows: $c=20$ yields a similar number of values to maintain in memory for the count matrix  based algorithm as the Hermite series based algorithm with $N=20$, i.e., this constitutes a like-for-like comparison. In \cite{xiao2019novel} a rule of thumb of $c=30$ cut-points is suggested for the count matrix based algorithm to give good performance for estimating the Spearman correlation coefficient. Larger numbers of cut-points are expected to give even better performance. We limit the maximum cut-points to 30 however since this already implies storing $961$ values for the count matrix versus the Hermite series algorithm where the maximum number of values to store is 483 for $N=20$. It is also worth noting that the execution times of the Hermite series based and count matrix based algorithms are roughly comparable (in our implementation). The results are summarized as the average MAE (and standard deviation of MAE) across all the values of $\rho$ in Table \ref{tableStat3} for the Hermite series based algorithm with $N=20$ and count matrix algorithm with $c=20$. The results for the Hermite series based algorithm with $N=20$ and count matrix algorithm with $c=30$ are presented in Table \ref{tableStat4}.
\begin{table}[t]
\caption{Summarized MAE results for Hermite series based Spearman rank correlation estimator ($N=20$) versus count matrix based Spearman rank correlation estimator ($c=20$) at $\mu=(0,0)$ for $\sigma_{1}=1, \sigma_{2}=1$ across all values of $\rho$. Lowest average MAE values for a given $n$ are presented in bold.}
\label{tableStat3}
\small
\centering
\begin{tabular}{>{}c>{}c>{}c>{}cc}
\hline
\multicolumn{1}{c}{Sample Size (n)} & \multicolumn{2}{c}{Average MAE (x10\textasciicircum{}-2)} & \multicolumn{2}{c}{Standard Deviation MAE (x10\textasciicircum{}-2)} \\
\cline{1-1} \cline{2-3} \cline{4-5}
 & Hermite (N = 20) & Matrix (c = 20) & Hermite (N = 20) & Matrix (c = 20)\\
10,000 & \textbf{0.180} & 0.193 & 0.006 & 0.083\\
50,000 & \textbf{0.079} & 0.188 & 0.002 & 0.094\\
100,000 & \textbf{0.055} & 0.187 & 0.001 & 0.096\\
\hline
\end{tabular}
\end{table}
\begin{table}[t]
\caption{Summarized MAE results for Hermite series based Spearman rank correlation estimator ($N=20$) versus count matrix based Spearman rank correlation estimator ($c=30$) at $\mu=(0,0)$ for $\sigma_{1}=1, \sigma_{2}=1$ across all values of $\rho$. Lowest average MAE values for a given $n$ are presented in bold.}
\label{tableStat4}
\small
\centering
\begin{tabular}{>{}c>{}c>{}c>{}cc}
\hline
\multicolumn{1}{c}{Sample Size (n)} & \multicolumn{2}{c}{Average MAE (x10\textasciicircum{}-2)} & \multicolumn{2}{c}{Standard Deviation MAE (x10\textasciicircum{}-2)} \\
\cline{1-1} \cline{2-3} \cline{4-5}
 & Hermite (N = 20) & Matrix (c = 30) & Hermite (N = 20) & Matrix (c = 30)\\
10,000 & 0.180 & \textbf{0.091} & 0.006 & 0.034\\
50,000 & \textbf{0.079} & 0.087 & 0.002 & 0.041\\
100,000 & \textbf{0.055} & 0.087 & 0.001 & 0.042\\
\hline
\end{tabular}
\end{table}
These results suggest that the Hermite series based algorithm performs better than the count matrix algorithm as the sample size increases which is particularly relevant in the streaming data and massive data set scenarios. We have repeated the same simulation analysis for values of $\mu$ away from the origin. We have observed that while both the Hermite series based and count matrix based Spearman rank correlation estimation algorithms perform well for the mean vector at the origin, performance deteriorates for both algorithms with mean-vectors away from the origin. The degradation in performance suggests that both algorithms would benefit from standardizing the observations. Thus, we implemented an online standardization procedure and applied it identically for both algorithms (see appendix A of \cite{stephanou2017sequential} for a description of calculating the mean and standard deviation in a stable, online manner). Comfortingly, the results for different values of $\mu$ and $\sigma_{1},\sigma_{2}$ are very similar to Table \ref{tableStat3} and \ref{tableStat4} with the online standardization procedure in place.

Next we simulate from a non-normal bivariate distribution by first drawing i.i.d. observations from a standard bivariate normal distribution, $(x_{i},y_{i}), \, i \in \{1,\dots,n\}$ and then transforming these observations with a strictly monotonically increasing function, $g(x)$, as $(g(x_{i}),g(y_{i})), \, i \in \{1,\dots,n\}$. Since $g(x)$ preserves orderings of $x_{i}$ and $y_{i}$ individually, the Spearman correlation coefficient will be unchanged. We choose $g(x)=\exp(x)$. The marginal distributions of $x$ and $y$ are then log-normal distributions. The results are summarized in Table \ref{tableStat5} and \ref{tableStat6}. We see that the performance of the Hermite series based algorithm is better than the count matrix algorithm with $c=20$ (a like-for-like comparison) and somewhat worse compared to the count matrix algorithm with $c=30$.
\begin{table}[t]
\caption{Summarized MAE results for Hermite series based Spearman rank correlation estimator ($N=20$) versus count matrix based Spearman rank correlation estimator ($c=20$) for bivariate normal variables transformed as $(g(x_{i}),g(y_{i}))$ with $g(x)=\exp(x)$, across all values of $\rho$. Lowest average MAE values for a given $n$ are presented in bold.}
\label{tableStat5}
\small
\centering
\begin{tabular}{>{}c>{}c>{}c>{}cc}
\hline
\multicolumn{1}{c}{Sample Size (n)} & \multicolumn{2}{c}{Average MAE (x10\textasciicircum{}-2)} & \multicolumn{2}{c}{Standard Deviation MAE (x10\textasciicircum{}-2)} \\
\cline{1-1} \cline{2-3} \cline{4-5}
 & Hermite (N = 20) & Matrix (c = 20) & Hermite (N = 20) & Matrix (c = 20)\\
10,000 & \textbf{0.897} & 1.072 & 0.527 & 0.592\\
50,000 & \textbf{0.812} & 0.939 & 0.634 & 0.547\\
100,000 & \textbf{0.810} & 0.909 & 0.675 & 0.540\\
\hline
\end{tabular}
\end{table}
\begin{table}[t]
\caption{Summarized MAE results for Hermite series based Spearman rank correlation estimator ($N=20$) versus count matrix based Spearman rank correlation estimator ($c=30$) for bivariate normal variables transformed as $(g(x_{i}),g(y_{i}))$ with $g(x)=\exp(x)$, across all values of $\rho$. Lowest average MAE values for a given $n$ are presented in bold.}
\label{tableStat6}
\small
\centering
\begin{tabular}{>{}c>{}c>{}c>{}cc}
\hline
\multicolumn{1}{c}{Sample Size (n)} & \multicolumn{2}{c}{Average MAE (x10\textasciicircum{}-2)} & \multicolumn{2}{c}{Standard Deviation MAE (x10\textasciicircum{}-2)} \\
\cline{1-1} \cline{2-3} \cline{4-5}
 & Hermite (N = 20) & Matrix (c = 30) & Hermite (N = 20) & Matrix (c = 30)\\
10,000 & 0.897 & \textbf{0.690} & 0.527 & 0.361\\
50,000 & 0.812 & \textbf{0.549} & 0.634 & 0.300\\
100,000 & 0.810 & \textbf{0.517} & 0.675 & 0.289\\
\hline
\end{tabular}
\end{table}

\subsection{Non-Stationary Data} \label{simnonstat}

\begin{figure}[h]
    \centering
    \includegraphics[width=65mm]{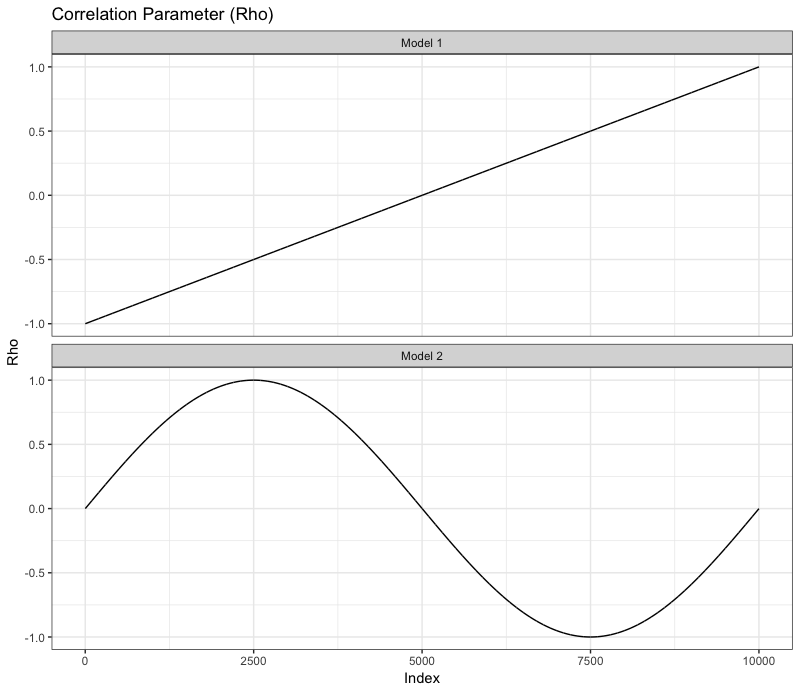}
    \caption{Evolution of the correlation parameter $\rho$ of the bivariate normal distribution used for the non-stationary models. Model 1 produces a stream of observations with monotonically increasing correlation. Model 2 produces a stream of observations that oscillates between correlated and anti-correlated regimes.}
    \label{fig_models}
\end{figure}

\begin{figure}[h]
    \centering
    \includegraphics[width=100mm]{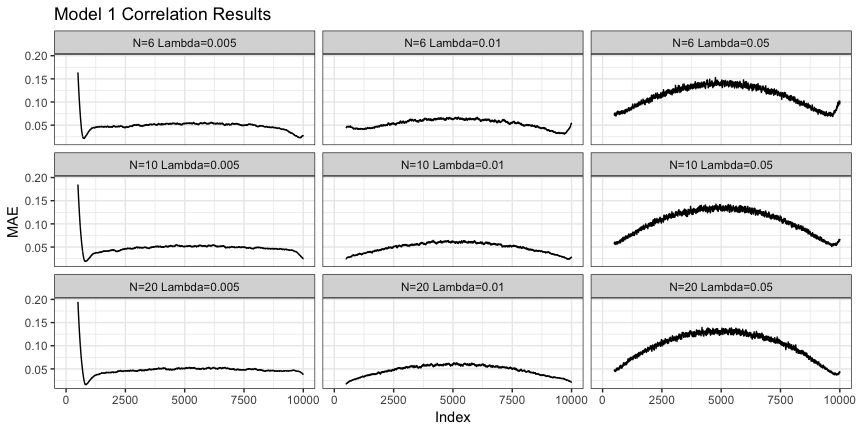}
     \caption{Mean absolute error results for the Hermite series based Spearman rank correlation estimator for Model 1 on a grid of $N$ and $\lambda$ values. These are the parameters controlling the behaviour of the estimator as discussed in Section \ref{lam_N_selection}. In particular, we evaluate $N \in \{6,10,20\}$ and $\lambda \in \{0.005,0.01,0.05\}$.}
    \label{fig_model_1_res_spear}
\end{figure}

\begin{figure}[h]
    \centering
    \includegraphics[width=110mm]{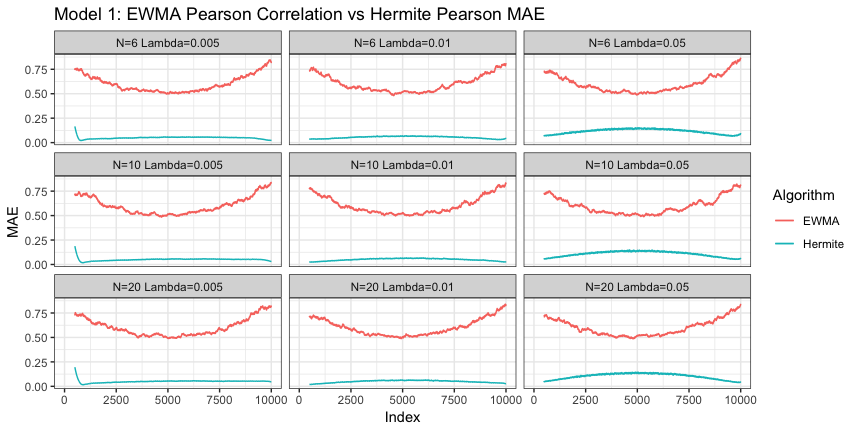}
    \caption{Mean absolute error results for the Hermite series based Spearman rank correlation estimator transformed to give Pearson's product-moment correlation compared to an exponentially weighted version of the standard Pearson's product-moment correlation estimator ($\lambda$ is the same for both algorithms).}
    \label{fig_model_1_res_pearson}
\end{figure}

\begin{figure}[h]
    \centering
    \includegraphics[width=100mm]{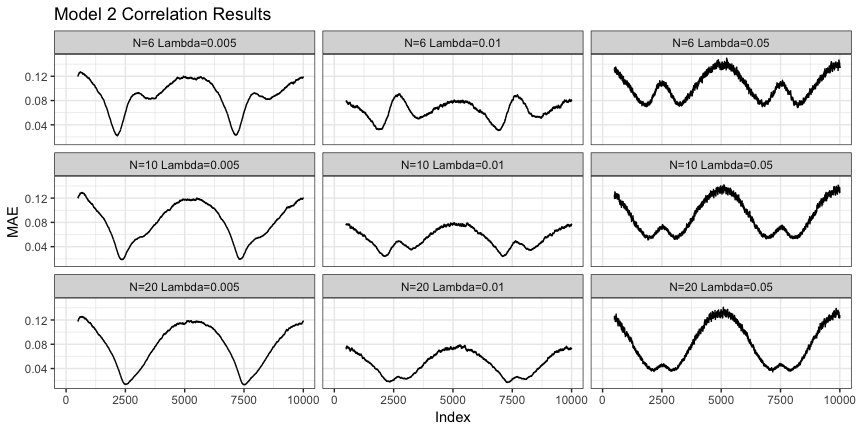}
    \caption{Mean absolute error results for the Hermite series based Spearman rank correlation estimator for Model 2 on a grid of $N$ and $\lambda$ values. These are the parameters controlling the behaviour of the estimator as discussed in Section \ref{lam_N_selection}. In particular, we evaluate $N \in \{6,10,20\}$ and $\lambda \in \{0.005,0.01,0.05\}$.}
    \label{fig_model_2_res_spear}
\end{figure}

\begin{figure}[h]
    \centering
    \includegraphics[width=110mm]{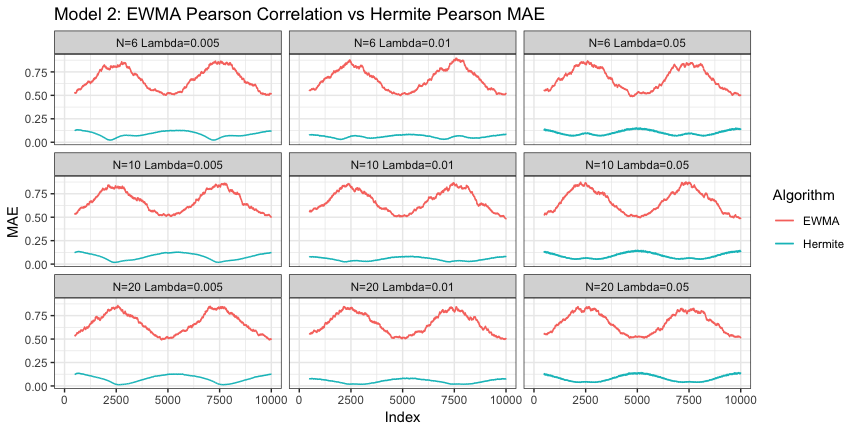}
      \caption{Mean absolute error results for the Hermite series based Spearman rank correlation estimator transformed to give Pearson's product-moment correlation compared to an exponentially weighted version of the standard Pearson's product-moment correlation estimator ($\lambda$ is the same for both algorithms).}
    \label{fig_model_2_res_pearson}
\end{figure}

In this section, we evaluate our proposed online algorithm applicable to non-stationary streams. As discussed in the previous section, the count matrix approach is not directly comparable in this setting. As such we do not include comparisons to the count matrix approach in this section. The non-stationary models we evaluate are the following: 
\begin{enumerate}
	\item We draw $n=10^{4}$ observations from a bivariate normal distribution with mean vector $\mu = (0,0)$ and covariance matrix  $\Sigma = (1, \rho^{(i)}; \rho^{(i)},1)$, where $\rho^{(i)} = -1 + 2(i-1)/(n-1)$, $i \in \{1,\dots,n\}$. Thus the correlation begins at $-1$ and ends at $1$. This models a bivariate stream that begins perfectly anti-correlated and ends perfectly correlated. In addition, we replace $0.5 \%$, i.e., $50$ observations uniformly at random with gross errors modelled by a bivariate normal with mean vector $\mu = (0,0)$ and covariance matrix $\Sigma = (10^{4}, 0; 0,10^{4})$. This allows us to also explore the robustness of the Hermite series based Spearman rank correlation estimation algorithm;
	\item We draw $n=10^{4}$ observations from a bivariate normal distribution with mean vector $\mu = (0,0)$ and covariance matrix  $\Sigma = (1, \rho^{(i)}; \rho^{(i)},1)$, where $\rho^{(i)}  = \sin(2\pi (i-1)/(n-1))$, $i \in \{1,\dots,n\}$. This models a bivariate stream that oscillates between correlated and anti-correlated regimes. This could represent the price return innovations of two financial time series which switch between momentum and mean-reversion regimes for example. In addition, we replace $0.5 \%$, i.e., $50$ observations uniformly at random with gross errors modelled by a bivariate normal with mean vector $\mu = (0,0)$, $\Sigma = (10^4, 0; 0,10^4)$.
\end{enumerate}
In Fig. \ref{fig_models} we plot the evolution of the correlation parameter $\rho$ for the two models described above.
To assess the robustness of the exponentially weighted Hermite series based Spearman rank correlation estimator, we also transform the Spearman rank correlation to Pearson's product-moment correlation (using the Fisher consistent estimator, $\hat{\rho}_{S} = 2 \sin \left( (\pi/6)\hat{R}_{N}\right)$,  applicable to bivariate normal distributions) and compare to an online, exponentially weighted version of the standard Pearson's correlation estimator, $\hat{\rho}(x,y)_{\lambda}$, which we define in \ref{ewpearson}. While the Spearman rank based estimator for the Pearson's product-moment correlation is not expected to be as efficient as the standard Pearson's correlation estimator, it allows us to assess robustness and accuracy in a comparable manner. For each of the two models, the following steps are repeated:
\begin{enumerate}
	\item Draw $n=10^{4}$ observations from the model;
	\item  Iterate through the sample, updating the exponentially weighted Hermite series based Spearman rank correlation estimate in accordance with algorithm 2 in Section \ref{nonstationaryAlgo} at each observation, $i$. Note that we did not apply an online standardization procedure to the observations. In addition, the Pearson's correlation is estimated from the Hermite series based Spearman rank estimate as $\hat{\rho}^{(i)}_{S} = 2 \sin \left( (\pi/6)\hat{R}^{(i)}_{N}\right)$; 
	\item Calculate the exact, large sample limit of the Spearman correlation coefficient (i.e. the grade correlation) at each observation through $R^{(i)}=(6 / \pi)\arcsin(\rho^{(i)} /2)$, where $\rho^{(i)} $ is the exact Pearson's correlation at each observation generated by the model under consideration;
	\item Repeat the aforementioned steps $m=10^3$ times and estimate two MAE curves. The first MAE curve tracks the MAE at the $i$th observation between the Hermite series based Spearman correlation estimate and the exact, large sample limit of the Spearman rank coefficient. The second MAE curve tracks the MAE at the $i$th observation between $\hat{\rho}^{(i)}_{S} = 2 \sin \left( (\pi / 6)\hat{R}^{(i)}_{N}\right)$, and the exact Pearson's correlation. The latter MAE curve is compared to the MAE performance of an exponentially weighted version of the standard Pearson's correlation estimator;
	\item Repeat this procedure at a grid of $N$ and $\lambda$ values to assess the performance of the Hermite series based estimator for different values of these parameters. Given that the selection of these parameters requires balancing several considerations, we assess performance on a grid of parameters to evaluate sensitivity of the results to the parameter choice. In addition, we seek to determine whether smaller values of $\lambda$ perform better as discussed in Section \ref{lam_N_selection}.  
\end{enumerate}
The Spearman rank correlation results for model 1 are presented in Fig. \ref{fig_model_1_res_spear}. Note that the results for model 1 excluding outliers (omitted for brevity) are very similar to the results with outliers as presented in Fig. \ref{fig_model_1_res_spear}, providing good evidence for the robustness of the Hermite series based estimator. The Pearson's correlation results for model 1 are presented in Fig. \ref{fig_model_1_res_pearson}.
The best results are achieved for $\lambda=0.005$ and the MAE is relatively insensitive to the choice of $N$.
\begin{figure}[h]
    \centering
    \includegraphics[width=110mm]{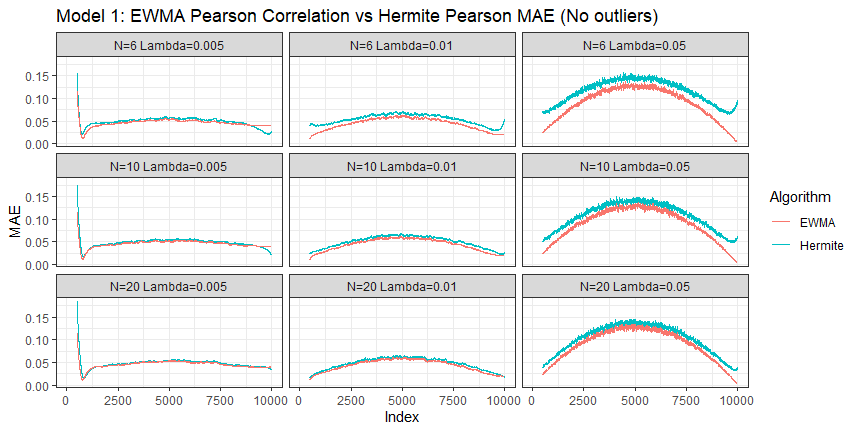}
    \caption{Mean absolute error results for the Hermite series based Spearman rank correlation estimator transformed to give Pearson's product-moment correlation compared to an exponentially weighted version of the standard Pearson's product-moment correlation estimator ($\lambda$ is the same for both algorithms) for model 1 with no outliers.}
    \label{fig_model_1_res_pearson_no_out}
\end{figure}
It is clear that the Pearson's correlation derived from the Hermite series based Spearman rank correlation estimator is more accurate and robust than an exponentially weighted version of the standard Pearson's product-moment correlation estimator. This is to be contrasted with the case without outliers in Fig. \ref{fig_model_1_res_pearson_no_out}, where the performance of the two estimators is very similar. The Spearman rank correlation results for model 2 are presented in Fig. \ref{fig_model_2_res_spear}. Again the results for model 2 excluding outliers (omitted for brevity) are very similar to the results with outliers as presented in Fig. \ref{fig_model_2_res_spear}. The Pearson's product-moment results for model 2 are presented in Fig. \ref{fig_model_2_res_pearson}.
The best results are achieved for $\lambda=0.01$ with higher values of $N$ performing slightly better than lower values ($N=20$ yields the best performance).
\begin{figure}[h]
    \centering
    \includegraphics[width=110mm]{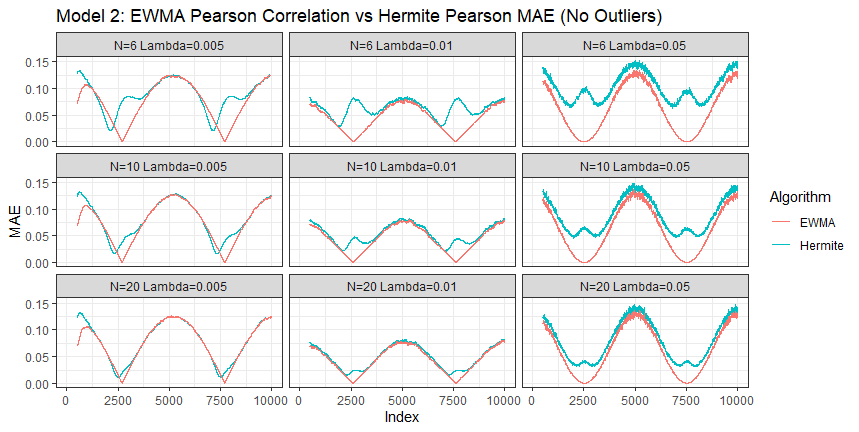}
      \caption{Mean absolute error results for the Hermite series based Spearman rank correlation estimator transformed to give Pearson's product-moment correlation compared to an exponentially weighted version of the standard Pearson's product-moment correlation estimator ($\lambda$ is the same for both algorithms) for model 2 with no outliers..}
    \label{fig_model_2_res_pearson_no_out}
\end{figure}
The Pearson's correlation derived from the Hermite series based Spearman rank correlation estimator is again more accurate and robust than an exponentially weighted version of the standard Pearson's product-moment correlation estimator. The performance of the two estimators is very similar in the case with no outliers as can be seen in Fig. \ref{fig_model_2_res_pearson_no_out}. 

\section{Real Data Example}\label{realResults}

In this section, we assess the exponentially weighted Hermite series based Spearman rank correlation estimator applied to tick-by-tick forex data (sourced from \cite{truefx}). The association between two currency pairs (EURUSD and GBPUSD in this instance) is expected to vary over time and is thus a good example of a non-stationary setting.

\subsection{Data Description}

The forex data is sourced from \cite{truefx} and is comprised of tick-by-tick, top-of-book bid and offer quotes (aggregated across several bank liquidity providers) for EURUSD and GBPUSD for April 2019. The mid-price series for EURUSD, $p^{(1)}_{i}$, and GBPUSD,  $p^{(2)}_{j}$, are calculated from the bid and offer quotes as follows:
\begin{equation*}
p^{(1)}_{i} = (b^{(1)}_{i} + a^{(1)}_{i})/2 \quad i \in \{1,\dots,n_{1}\}, \quad
p^{(2)}_{j} = (b^{(2)}_{j} + a^{(2)}_{j})/2 \quad j \in \{1,\dots,n_{2}\},
\end{equation*}
where $a^{(1)}, a^{(2)}$ are EURUSD and GBPUSD offer quotes respectively and $b^{(1)}, b^{(2)}$ are EURUSD and GBPUSD bid quotes respectively. These mid-prices are then sampled on a minutely basis using the previous-tick methodology (i.e. the most recent tick in each currency pair is recorded as the price at a given minute), see \cite{ait2010high} for example for a description of this methodology in the context of other approaches to data synchronization. The mid-price minutely sampled series are thus:
\begin{equation*}
p^{(1)}_{t_{(1)}},\dots, p^{(1)}_{t_{(n)}},\quad
p^{(2)}_{t_{(1)}},\dots, p^{(2)}_{t_{(n)}} , \quad t_{(i+1)}-t_{(i)} = 1\mbox{ minute}, \quad i \in \{1,\dots,n\},
\end{equation*}
and we obtain the basis point log returns returns via:
\begin{equation}
\label{basisptret}
r^{(1)}_{(j)} =10^{4}\log \frac{p^{(1)}_{t_{(j+1)}}}{p^{(1)}_{t_{(j)}}}, \quad r^{(2)}_{(j)} =10^{4}\log \frac{p^{(2)}_{t_{(j+1)}}}{p^{(2)}_{t_{(j)}}}, \quad j \in \{1,\dots,n-1\}. 
\end{equation}
The descriptive statistics of the returns are presented in Table \ref{forex_descrip}. The data presents a mild violation of the assumption of a continuous bivariate distribution in that some values of basis point log returns are repeated (due to the fact that the minimum increment of quotes, i.e., tick size is 1e-5). Indeed $17.3\%$ of EURUSD log returns and $11\%$ of GBPUSD log returns are repeated.
\begin{table}[h]
\caption{Descriptive statistics of forex basis point log returns calculated as per equation \eqref{basisptret} for EURUSD and GBPUSD data for April 2019 sourced from  \cite{truefx}.}
\label{forex_descrip}
\centering
{\small
\begin{tabular}{ccc}
\hline
Summary Statistic & EURUSD Returns & GBPUSD Returns\\
\hline
Count & 30572.00 & 30572.00\\
Mean & 0.00 & 0.01\\
Standard Deviation & 0.77 & 1.21\\
Skewness & -0.66 & 0.08\\
Kurtosis & 21.06 & 26.53\\
Min & -17.50 & -26.67\\
Max & 7.49 & 21.97\\
\hline
\end{tabular}
}
\end{table}

\subsection{Results}

\begin{figure}[h]
    \centering
    \includegraphics[width=90mm]{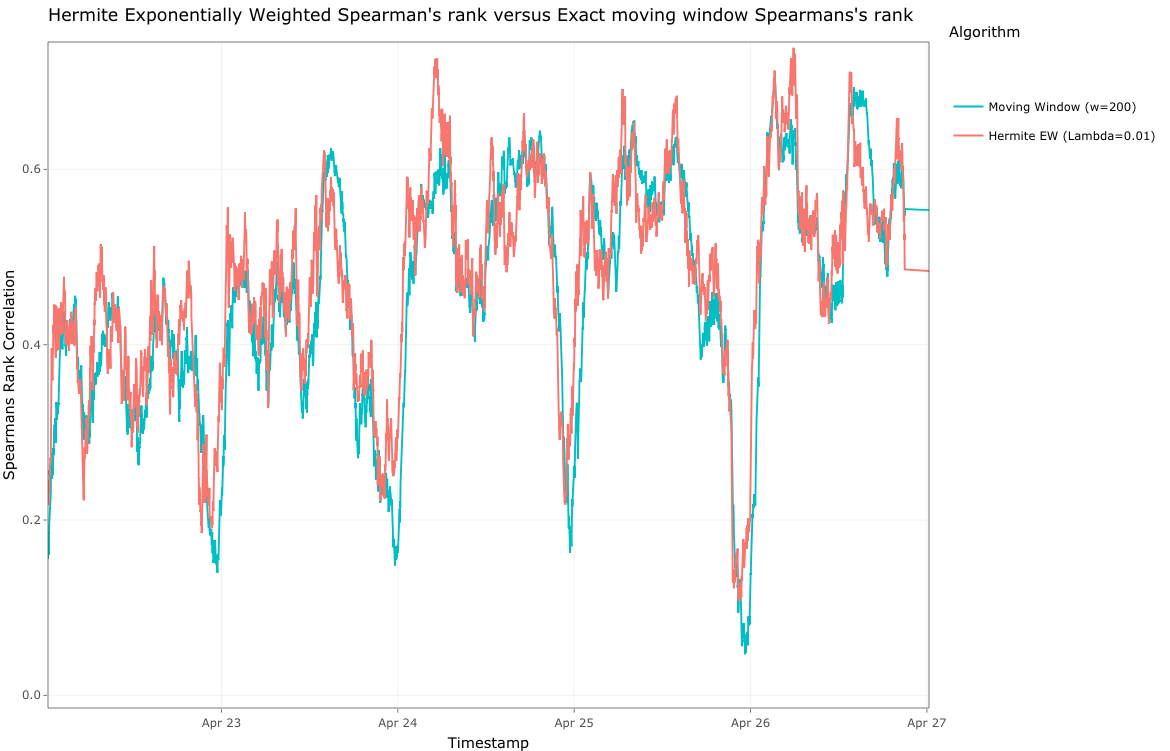}
      \caption{Hermite series based exponentially weighted Spearman rank estimator ($\lambda=0.01$) versus moving window exact Spearman correlation ($w=200$).}
    \label{fig_currency_results}
\end{figure}

In this section, we utilize our \textsf{R} code implementation which has been formalized in the package \texttt{hermiter} \cite{hermiter}. The initial analysis we conduct is as follows: we iterate through the bivariate forex returns data set described above and update the exponentially weighted Hermite series based Spearman rank correlation estimator at each minutely return update $(r^{(1)}_{(j)},r^{(2)}_{(j)}), \, j \in \{1,\dots,n - 1\}$. These estimates are then compared to exact Spearman rank correlation estimates, based on a trailing moving window of a certain size. In choosing the appropriate value for the exact Spearman rank correlation moving window size, $w$, we utilize the relation $\lambda = 2/(w+1)$ as described in Section \ref{lam_N_selection}. For analyzing the forex returns data at an average ``age"/lag of approximately two hours, we set $\lambda = 0.01$, corresponding to a moving window size $w=200$. Based on the simulation studies conducted previously, we have seen that the Hermite series based algorithm is  reasonably insensitive to the choice of $N$, we thus chose $N=10$ for illustrative purposes. The results are presented in Fig. \ref{fig_currency_results}.

It is clear that the exponentially weighted Hermite series based Spearman rank correlation estimator and the moving window exact Spearman rank correlation estimator track the association between the currency pairs similarly. Indeed the mean absolute difference between the estimates across all observations is $0.048$. 

In terms of forecasting ability, we predict the two hour forward realized Spearman rank correlation coefficient formed from observations $j+1, \dots, j+120$, where $j \in \{1000,\dots, (n-120)\}$. We have thus assumed approximate local stationarity of the data at the time scale of a few hours. The true value of the realized forward Spearman rank correlation is compared to the exact Spearman correlation estimate calculated on a trailing moving window, as well as to the exponentially weighted Hermite series based Spearman correlation estimate, both evaluated at observation $j$ where $j \in \{1000,\dots,(n-120)\}$. The latter estimates are the most recently computed Spearman correlation coefficients in the exponentially weighted scheme at each observation. To select the best $\lambda, N$ values for the Hermite series based estimator and the best $w$ value for the moving window estimator we analyzed forex data for the previous month of March 2019. In particular, we performed a grid-search for $\lambda$ and $N$, covering $\lambda \in \{0.001,0.002,0.005,0.01, 0.05\}$ and $N \in \{6,8,10,20\}$ for the exponentially weighted Hermite series Spearman rank correlation estimator. The best performing parameters were selected according to the MAE of the predicted Spearman rank correlation versus the realized forward Spearman rank correlation for the next two hours. These values were $\lambda=0.002, N=10$, respectively. Note that we again applied the online standardization procedure that we found advantageous in our simulation studies in Section \ref{simstudies_stat}. For the trailing moving window size of the exact Spearman correlation coefficient, we evaluated $w \in \{50,100,\dots,1000\}$. The best value for the moving window size was $w=850$. Even at $w=200$ the moving window approach was more computationally expensive and slower than the Hermite series based approach. We thus terminated the parameter search at $w=1000$. We also evaluated the constant prediction based on the batch estimate formed from calculating the Spearman rank correlation coefficient for the whole March 2019 data set. The out of sample results for the MAE of the estimates compared to the forward realized Spearman rank correlation coefficient in April 2019 are displayed in Table \ref{tableMAEreal}.
\begin{table}[t]
\caption{Out of sample MAE results for forecasting forward, realized two hour Spearman rank correlation.}
\label{tableMAEreal}
\small
\centering
\begin{tabular}{c c}
\hline
Algorithm & MAE\\ \hline
Hermite EW ($\lambda=0.002$) & 0.1103 \\ 
Moving Window ($w=850$) & 0.1120\\ 
Constant Estimate & 0.1211 \\ \hline
\end{tabular}
\end{table}

The Hermite series based exponentially weighted Spearman estimator is slightly more accurate than the exact Spearman correlation evaluated on a trailing moving window and significantly more accurate than the batch estimate formed from March 2019. The main advantages of the Hermite series based exponentially weighted estimator are as follows. Firstly, the Hermite series based Spearman rank correlation estimator has $O(1)$ time and memory complexity with respect to $\lambda$. By contrast, the trailing moving window approach grows in time complexity and memory requirements with moving window size. For large moving window sizes with the usual Spearman correlation estimator, the Hermite series based estimator will be significantly faster. As an example, with our implementation in \textsf{R} \cite{hermiter}, the moving window estimator (using cor in \textsf{R}) with $w=1000$ is roughly 3 times slower than the Hermite series based algorithm and this differential increases rapidly with $w$. Secondly, observations older than a certain threshold do not sharply drop off, but are rather incorporated with decreasing weighting.

\section{Conclusion and Perspectives}\label{conclusion}

In this article we have introduced novel, sequential (online) algorithms to estimate the most popular measure of nonparametric correlation, the Spearman rank correlation coefficient, in both stationary (static) and non-stationary (dynamic) settings. These algorithms are based on bivariate Hermite series density estimators and Hermite series based distribution function estimators. The stationary setting corresponds to bivariate data streams where the Spearman rank correlation coefficient is constant. This setting also includes the one-pass  estimation of the Spearman rank correlation for massive data sets. In addition, this algorithm could be applied in decentralized settings where Hermite series based estimates of the Spearman rank correlation, generated on separate portions of a larger dataset, can be combined to form an overall estimate. We have proved that in the i.i.d. setting, the Hermite series based Spearman rank correlation estimator converges to the grade correlation which is the constant that is unbiasedly estimated by the standard Spearman rank correlation estimator in large samples. The non-stationary (dynamic) setting corresponds to situations where the Spearman rank correlation of a bivariate data stream is expected to vary over time. In order to treat these scenarios we have introduced an exponentially weighted version of the Hermite series based Spearman rank correlation estimator. To the best of our knowledge this is the first algorithm allowing the online tracking of a time varying Spearman rank correlation of a bivariate data stream that does not rely on moving/sliding windows. We have derived variance (and standard error) results for the exponentially weighted estimator in the i.i.d. scenario and used these results to provide rule of thumb guidance for the selection of the weighting parameter, $\lambda$. The effectiveness of the Hermite series based Spearman rank estimator has been demonstrated in the stationary setting by means of a simulation study which revealed the Hermite series based estimator to be competitive with an existing approach. We have also demonstrated the effectiveness of the exponentially weighted Hermite estimator through simulation and real data studies. We expect our algorithms to be useful in a variety of settings, particularly those where robustness to outliers and errors is required in addition to fast online calculation of correlation which may vary over time (financial applications involving high frequency data streams for example). The same approach can also be applied to define a Hermite series based Kendall rank correlation estimator. We hope to report on these developments in the near future.

\subsection{Future Directions}

Some directions for future work include extending and improving the theoretical results presented in this article and exploring applications in machine learning. For example, in Theorem \ref{mainTheoremMAE}, the regularity conditions on $f(x,y)$ and associated marginal densities imply a fairly high degree of smoothness. It may be possible to weaken these conditions however, further generalizing the result. In addition, it is possible that a better rate of convergence could be derived. Potential applications in the machine learning domain include, firstly, feature selection on massive datasets. In particular, a univariate filter approach based on our Spearman rank correlation estimator can be applied, which is more general and robust than the Pearson's correlation, and could be a more computationally efficient alternative to information gain (mutual information). Another potential application pertains to hierarchical clustering on massive data sets or streaming data. A dissimilarity matrix based on Spearman correlation can be calculated in real time (for streaming data), or in one pass (for massive data sets), leveraging the  computational advantages of the Hermite series based Spearman rank correlation estimator. This dissimilarity matrix can then be used along with standard agglomerative or divisive hierarchical clustering algorithms. For a reasonably small number of variables this should facilitate real-time hierarchical clustering. This is again a potentially fruitful area for future research.

\section*{Acknowledgments}

We would like to sincerely thank the referees, the Associate Editor and especially the Editor. The insightful and very useful comments and suggestions we received greatly helped us to improve this article.

\appendix

\section{Proofs} \label{appendix_proofs}
In the proof of Theorem \ref{mainTheoremMAE}, we make use of the following two lemmas proved below.

\begin{lemma} \label{lemmaMaxVal}
For the Hermite series based distribution function estimators of the marginal cumulative distribution functions $F^{(1)}(x)$ and $F^{(2)}(y)$, we have,
\begin{equation*}
\max_{x}(|\hat{F}_{N}^{(1)}(x)|)  = O(N^{17/12}), \qquad
\max_{y}(|\hat{F}_{N}^{(2)}(y)|) = O(N^{17/12}).
\end{equation*}
\end{lemma}

\begin{proof}[\textbf{\upshape Proof of Lemma \ref{lemmaMaxVal}:}]
Using \eqref{HermiteSeriesDistroEst} and lemma 4 of \cite{stephanou2020properties},
\begin{equation}
 |\hat{F}_{N}^{(1)}(x)| \leq \sum_{k=0}^{N} |\hat{a}_k| \int_{-\infty}^{x}|h_{k}(t)| dt \leq u_{1}\sum_{k=0}^{N} (k+1)^{-\frac{1}{3}} + v_{1} \sum_{k=0}^{N}(k+1)^{\frac{5}{12}} = O(N^{17/12}),\nonumber 
\end{equation}
where $u_{1}$ and $v_{1}$ are positive constants. Note that we have used the fact that $\max_{x} |h_{k}(x)| \leq C (k+1)^{-1/12}$ where $C$ is a positive constant (implied by Theorem 8.91.3 of \cite{szeg1939orthogonal}) which implies $\max_{x}|\hat{a}_k| \leq C (k+1)^{-1/12}$. Similarly $\max_{y}(|\hat{F}_{N}^{(2)}(y)|) =O(N^{17/12})$.
\end{proof}

\begin{lemma} \label{lemmaMultivariateMIAE}
Suppose $f(x,y) \in L_2$ and $(x-\partial_{x})^{r_3}(y-\partial_{y})^{r_3}f(x,y) \in L_{2}$, $r_3 > 2$ fixed and finite. In addition, suppose $\mbox{E}|XY|^{2/3} < \infty$ and that $N_{3}(n) = N_{4}(n) = O(n^{1/(2r_3+1)})$ as $n \to \infty$, then we have,
\begin{equation}
\mbox{E}\int|\hat{f}_{N_{3}N_{4}}(x,y)-f(x,y)|dxdy = O(n^{-(r_3-2)/(2r_3+1)}) = o(1) \mbox{ uniformly in x and y},\nonumber 
\end{equation}
where $\hat{f}_{N_{3}N_{4}}(x,y)$ is the bivariate Hermite series estimator \eqref{HermiteSeriesProbEstBivariate}.
\end{lemma}

\begin{proof}[\textbf{\upshape Proof of Lemma \ref{lemmaMultivariateMIAE}:}]
From the inequalities implied by Theorem 8.91.3 of \cite{szeg1939orthogonal}, namely, 
\begin{equation}
\max_{|x|\leq a} h_{k}(x) \leq c_{a} (k+1)^{-\frac{1}{4}}, \qquad \max_{|x|\geq a}|h_{k}(x)|x^{\lambda} \leq d_{a} (k+1)^{s}, \label{outdomainInequal}
\end{equation}
where $c_{a}$ and $d_{a}$ are constants, $s=\max(\lambda / 2 - 1/12, -1/4)$, we have:
\begin{equation*}
\mbox{E}(\hat{A}_{kl}-A_{kl})^{2} = \frac{1}{n} \mbox{Var}(h_{k}(x) h_{l}(y)) \leq \frac{1}{n} \mbox{E}\left[(h_{k}(x))^{2} (h_{l}(y))^2\right] \leq \frac{c}{n} \mbox{E}(|XY|^{2/3}) (k+1)^{-1/2}(l+1)^{-1/2}, 
\end{equation*}
where we have set $\lambda=-1/3$ in the inequality \eqref{outdomainInequal} and $c$ is a positive constant. Note that this is satisfied when $\mbox{E}|X|^{4/3} < \infty$ and $\mbox{E}|Y|^{4/3} < \infty$ by the Cauchy-Schwarz inequality. Now using the definition of \eqref{HermiteSeriesProbEstBivariate} along with the monotone convergence theorem and the Lyapunov inequality we have,
\begin{align*}
\mbox{E}&\int|\hat{f}_{N_{3}N_{4}}(x,y)-f(x,y)|dxdy \MoveEqLeft[1]\\\
&\leq \sum_{k=0}^{N_{3}}\sum_{l=0}^{N_{4}} \mbox{E}|A_{kl} - \hat{A}_{kl}| \int_{-\infty}^{\infty}|h_k(x)|dx \int_{-\infty}^{\infty}|h_l(y)|dy
+ \sum_{k=N_{3}+1}^{\infty}\sum_{l=N_{4}+1}^{\infty} |A_{kl}|  \int_{-\infty}^{\infty}|h_k(x)|dx \int_{-\infty}^{\infty}|h_l(y)|dy\\
&\leq \sum_{k=0}^{N_{3}}\sum_{l=0}^{N_{4}} \sqrt{\mbox{E}|A_{kl} - \hat{A}_{kl}|^2} \int_{-\infty}^{\infty}|h_k(x)|dx \int_{-\infty}^{\infty}|h_l(y)|dy
+ \sum_{k=N_{3}+1}^{\infty}\sum_{l=N_{4}+1}^{\infty} |A_{kl}|  \int_{-\infty}^{\infty}|h_k(x)|dx \int_{-\infty}^{\infty}|h_l(y)|dy.
\end{align*}
In a similar way to \cite{convergence1}, if $(x-\partial_{x})^{r_3}(y-\partial_{y})^{r_3}f(x,y) \in L_{2}$, then,
\begin{equation}
|A_{kl}| \leq |B_{kl}| 2^{-r_3} (k+1)^{-r_{3}/2} (l+1)^{-r_{3}/2}, \label{biasBivar}
\end{equation}
where $B_{kl}$ are the bivariate Hermite expansion coefficients of $(x-\partial_{x})^{r_3}(y-\partial_{y})^{r_3}f(x,y)$. Note that this result relies on an asymptotic expansion of the Hurwitz zeta function and only applies for $r_3$ fixed and finite (see \citep{nemes2017error}, equation 1.3 for the form of the asymptotic expansion). Utilizing the results for the variance of $\hat{A}_{kl}$ above along with lemma 4 of \cite{stephanou2020properties}, \eqref{biasBivar} and the Cauchy-Schwarz inequality we have,
\begin{equation*}
 \mbox{E}\int|\hat{f}_{N_{3}N_{4}}(x,y)-f(x,y)|dxdy =O\left(\frac{N_{3}^{5/4} N_{4}^{5/4}}{\sqrt{n}}\right) + O\left(N_{3}^{-r_{3}/2 +1} N_{4}^{-r_{3}/2 +1}\right).
\end{equation*}
Now if we set $N_{3}(n) = N_{4}(n) = O(n^{1/(2r_3+1)})$, we obtain:
\begin{equation*}
\mbox{E}\int|\hat{f}_{N_{3}N_{4}}(x,y)-f(x,y)|dxdy = O(n^{-(r_3-2)/(2r_3+1)}).
\end{equation*}
\end{proof}

\begin{proof}[\textbf{\upshape Proof of Theorem \ref{mainTheoremMAE}:}]
We utilize \eqref{gradecorrel} and   \eqref{hermiteSpearmansEst} along with the Cauchy-Schwarz inequality, lemma \ref{lemmaMaxVal} and Jensen's inequality to obtain,
\begin{align}
\mbox{E}&|\hat{R}_{N_{1},N_{2},N_{3},N_{4}} -  \rho(F^{(1)}(X), F^{(2)}(Y))| \MoveEqLeft[1]\nonumber\\ 
\leq  &O\left((N_{1}N_{2})^{17/12}\right) \mbox{E}\int|\hat{f}_{N_{3}N_{4}}(x,y)-f(x,y)|dxdy + O\left(N_{1}^{17/12}\right)  \sqrt{\mbox{E}\int (\hat{F}_{N_{2}}^{(2)}(y)-F^{(2)}(y))^2 f(y)dy}  \nonumber\\
&+ 12\sqrt{\mbox{E}\int (\hat{F}_{N_{1}}^{(1)}(x)-F^{(1)}(x))^2 f(x)dx} +  O\left(N_{1}^{17/12}\right) \mbox{E}\int |\hat{f}_{N_{3}N_{4}}(x,y)-f(x,y)| dxdy \nonumber \\
&+  6 \sqrt{\mbox{E} \int (\hat{F}_{N_{1}}^{(1)}(x) - F^{(1)}(x))^2 f(x)dx} +  O\left(N_{2}^{17/12}\right)  \mbox{E}\int|\hat{f}_{N_{3}N_{4}}(x,y)-f(x,y)| dxdy \nonumber\\ 
&+  6 \sqrt{\mbox{E}\int (\hat{F}_{N_{2}}^{(2)}(y) - F^{(2)}(y))^2 f(y)dy} +  3 \mbox{E}\int|\hat{f}_{N_{3}N_{4}}(x,y)-f(x,y)| dxdy. \nonumber
\end{align}
Now, given assumptions 1, 2 and 4 (also note that $\mbox{E}|X|^{4/3} < \infty$ and $\mbox{E}|Y|^{4/3} < \infty$ implies that $\mbox{E}|X|^{2/3} < \infty$ and $\mbox{E}|Y|^{2/3} < \infty$ by the Lyapunov inequality), Theorem 2 of \cite{stephanou2020properties} implies that,
\begin{equation*}
 \mbox{E} \int (\hat{F}_{N_{1}}^{(1)}(x) - F^{(1)}(x))^2 f(x)dx = O(N_{1}^{-r_1 +2}) + O(N_{1}^{5/2}/n) = O(n^{-2(r_1 -2)/(2r_1+1)}),
\end{equation*}
\begin{equation*}
 \mbox{E} \int (\hat{F}_{N_{2}}^{(2)}(y) - F^{(2)}(y))^2 f(y)dy = O(N_{2}^{-r_2 +2}) + O(N_{2}^{5/2}/n) = O(n^{-2(r_2 -2)/(2r_2+1)}).
\end{equation*}
Note that again these results rely on an asymptotic expansion of the Hurwitz zeta function and only apply for $r_1, r_2$ fixed and finite. Using the above results along with Lemma \ref{lemmaMultivariateMIAE} we have, for $n\to \infty$:
\begin{align*}
\mbox{E}|\hat{R}_{N_{1},N_{2},N_{3},N_{4}} -  \rho(F^{(1)}(X), F^{(2)}(Y))| &=  O(n^{34/12(2r_{1}+1)+34/12(2r_{2}+1)-(r_3-2)/(2r_3+1)}) \\
& +O(n^{34/12(2r_{1}+1) -(r_2 -2)/(2r_2+1)}) + O(n^{-(r_1 -2)/(2r_1+1)}).
\end{align*}
This implies that if we have $r_{1}, r_{2},r_{3} \geq 8$:
\begin{equation*}
\mbox{E}|\hat{R}_{N_{1},N_{2},N_{3},N_{4}} -  \rho(F^{(1)}(X), F^{(2)}(Y))| = o(1).
\end{equation*}
\end{proof}

\begin{proof}[\textbf{\upshape Proof of Theorem \ref{theoremVariance}:}]
We begin by noting that: 
\begin{equation}
	\mbox{Var}((\hat{a}_{(1)})_{r}) = \lambda^{2}\sum_{j=0}^{n-1} (1-\lambda)^{2j} \mbox{Var}(h_{r}(x_{n-j})) + (1-\lambda)^{2n}  \mbox{Var} (h_r(x_0)) = \left[ \frac{\lambda}{2-\lambda} (1-(1-\lambda)^{2n})+(1-\lambda)^{2n} \right] \mbox{Var} (h_r(x)). \nonumber
\end{equation}
Thus for $n \to \infty$, we have $\mbox{Var}((\hat{a}_{(1)})_{r}) \to (\lambda / (2-\lambda))\mbox{Var}(h_r(x))$ which is $O(\lambda)$. Similarly, for $n \to \infty$, we have that $\mbox{Var}((\hat{a}_{(2)})_{s}))$, $\mbox{Var}((\hat{A})_{uv})$, $\mbox{cov}((\hat{a}_{(1)})_{r},(\hat{a}_{(2)})_{s})$, $\mbox{cov}((\hat{a}_{(1)})_{r},(\hat{A}_{uv}))$, $\mbox{cov}((\hat{a}_{(2)})_{s},(\hat{A}_{uv}))$ are $O(\lambda)$. Therefore, all the variances and covariances above are $O(\lambda)$. An exact Taylor series expansion of $\hat{R}_{N}$ implies $\mbox{E}(\hat{R}_{N}) = \hat{R}_{N}|_{a_{(1)},a_{(2)},A} + O(\lambda)$ and facilitates a convenient reorganisation of the variance expression,  $\mbox{E}\left(\hat{R}_{N} - \mbox{E}(\hat{R}_{N}) \right)^{2}$. A straightforward but lengthy calculation yields the stated result for $n \to \infty$. Note that this essentially corresponds to the delta method as described in \cite{kendall1987kendall}.
\end{proof}

\section{Exponentially weighted Pearson's correlation Estimator} \label{ewpearson}

The exponentially weighted version of the standard Pearson's product-moment correlation estimator is defined as:
\begin{equation*}
	\bar{X}^{(1)}_{\lambda} =  \mathbf{x}_{1},\,
	\bar{X}^{(i)}_{\lambda} = \lambda \mathbf{x}_{i} + (1-\lambda)\bar{X}^{(i-1)}_{\lambda} , \quad
	\bar{Y}^{(1)}_{\lambda} =  \mathbf{y}_{1},\,
	\bar{Y}^{(i)}_{\lambda} = \lambda \mathbf{y}_{i} + (1-\lambda)\bar{Y}^{(i-1)}_{\lambda},
\end{equation*}
\begin{equation*}
	V(x)^{(1)}_{\lambda} =  1,\,
	V(x)^{(i)}_{\lambda} = \lambda (\mathbf{x}_{i} - \bar{X}^{(i)}_{\lambda})^{2}+ (1-\lambda)V(x)^{(i-1)}_{\lambda} , \quad
	V(y)^{(1)}_{\lambda} =  1,\,
	V(y)^{(i)}_{\lambda} = \lambda (\mathbf{y}_{i} - \bar{Y}^{(i)}_{\lambda})^{2}+ (1-\lambda)V(y)^{(i-1)}_{\lambda},
\end{equation*}
\begin{equation*}
	C(x,y)^{(1)}_{\lambda} =  1, \,
	C(x,y)^{(i)}_{\lambda} = \lambda (\mathbf{x}_{i} - \bar{X}^{(i)}_{\lambda}) (\mathbf{y}_{i} - \bar{Y}^{(i)}_{\lambda})+ (1-\lambda)C(x,y)^{(i-1)}_{\lambda} , \quad i \in \{2,\dots,n\},
\end{equation*}
\begin{equation*}
	\hat{\rho}(x,y)^{i}_{\lambda}  = \frac{C(x,y)^{(i)}_{\lambda}}{\sqrt{V(x)^{(i)}_{\lambda} V(y)^{(i)}_{\lambda} }}. 
\end{equation*}

\bibliographystyle{myjmva}
\bibliography{article}
\end{document}